\documentclass[aps,prl,twocolumn,showpacs,preprintnumbers]{revtex4}
\usepackage{amsmath,amssymb}
\usepackage{graphicx}
\usepackage{color}
\newcommand\beq{\begin{equation}}
\newcommand\eeq{\end{equation}}
\newcommand\beqa{\begin{eqnarray}}
\newcommand\eeqa{\end{eqnarray}}

\newcommand{\gs}{$\gamma_S$}
\newcommand{\ts}{$T_S$}
\newcommand{\mus}{$\mu_S$}
\newcommand{\vs}{$V_S$}
\newcommand{\tns}{$T_{NS}$}
\newcommand{\muns}{$\mu_{NS}$}
\newcommand{\vns}{$V_{NS}$}

\newcommand\etal{{\sl et al.\/}}
\newcommand\bl[1]{}

\newcommand\jhep{{J.\ H.\ E.\ P.\/}\ }
\newcommand\npa{{Nucl.\ Phys.\/} A\ }

\newcommand\plb{{Phys.\ Lett.\/} B\ }
\newcommand\zpc{{Z.\ Phys.\/} C\ }
\begin{document}

\title{Strange freezeout}
\author{S.\ \surname{Chatterjee}}
\email{sandeep@cts.iisc.ernet,in}
\affiliation{Center for High Energy Physics,\\ Indian Institute of Science,\\
         Bangalore 560012, India.}
\author{R.\ M.\ \surname{Godbole}}
\email{rohini@cts.iisc.ernet.in}
\affiliation{Center for High Energy Physics,\\ Indian Institute of Science,\\
         Bangalore 560012, India.}
\author{Sourendu \surname{Gupta}}
\email{sgupta@tifr.res.in}
\affiliation{Department of Theoretical Physics, Tata Institute of Fundamental
         Research,\\ Homi Bhabha Road, Mumbai 400005, India.}
\begin{abstract}
We argue that known systematics of hadron cross sections may cause
different particles to freeze out of the fireball produced in heavy-ion
collisions at different times.  We find that a simple model with two
freezeout points is a better description of data than that with a single
freezeout, while still remaining predictive. The resulting fits seem
to present constraints on the late stage evolution of the fireball,
including the tantalizing possibility that the QCD chiral transition
influences the yields at $\sqrt S=2700$ GeV and the QCD critical point
those at $\sqrt S=17.3$ GeV.
\end{abstract}
\maketitle

A fireball is formed in high-energy collision of heavy ions. Within
it scattering keeps particles in equilibrium. If the initial
collision energy is large enough then this matter could be formed
in the plasma phase of the strong interactions \cite{white}. This
fireball expands and cools. In the late stages of its evolution,
when inelastic scatterings are no longer frequent enough to maintain
chemical equilibrium between different hadron species, the system
is said to undergo chemical freezeout (CFO) \cite{heinz}. Ever since
the demonstration that data on yields of hadrons for a large range of
initial collision energies is explained by a single CFO temperature, $T$,
and baryon chemical potential $\mu$, this has been a paradigm for heavy
ion collisions \cite{andro,yen,cleymans}.  It has been clear that there
will be some corrections to this, since the full description of the CFO
process must involve kinetic equations, and therefore the densities
of hadrons and their reaction cross sections \cite{seq}. The expectation
was that these corrections are fairly small, and therefore substructure
in CFO is a detail. This paradigm is different from decoupling in the
early universe, where the hierarchy of interactions gave significantly
different CFO for photons and neutrinos.

However, in a hadronic fireball whose lifetime is short compared to the
time-scale of weak interactions, strangeness changing transmutations of
baryons occur in interactions with kaons, via intermediate states
of excited strange baryons. Furthermore, $\phi$ decays into pairs
of kaons 83\% of the time.  Although the width of $\phi$ is only 4.3
MeV, this is a reflection of reduced phase space volume rather than
a reduced interaction strength.  This means that the inverse reaction
has a low threshold, and, as a result, $\phi$ is in equilibrium with
the kaons. However, when kaons decouple, all these hadrons fall out of
equilibrium, so providing a natural mechanism for multiple CFO.

Since the pion density is high, and baryon isospin changing reaction
thresholds low, isospin transmutations can still proceed \cite{iso}.
Since baryons of different strangeness each remain connected to pions, they
have the same isospin chemical potential even when they are
otherwise chemically decoupled.  The relative yields of all isodoublets
are approximately equal because every sector communicates with a common
pion isospin bath.  It is impossible to test this simple prediction
because the doublets involve uncharged baryons, which are invisible
to detectors. Even the comparison of charged members of isotriplets,
$\Sigma^\pm$ and $\pi^\pm$, is ruled out because $\Sigma^\pm$ cannot
be reconstructed from its decays since the products include uncharged
hadrons.

\begin{table*}[bt]
\begin{center}
\begin{tabular}{|r@{.}l|c|c|c|c|c|c|c|c|c|}
\hline
 \multicolumn{2}{|c|}{$\sqrt{S_{NN}}$} & Ref &
 $10^4V_S$ & $10^4V_{NS}$ &
 $T_S$ & $T_{NS}$ &
 $\mu_S$ & $\mu_{NS}$ & $\chi^2/N_{df}$ \\
 \multicolumn{2}{|c|}{(GeV)} & &
 (MeV$^{-3}$) & (MeV$^{-3}$) & (MeV) & (MeV) & (MeV) & (MeV) & \\
\hline
   6&27  & \cite{na49a,na49b,na49c,na49d}
         &1.1 (0.2)&1.6 (0.3)&139 (4)&131 (4)&435 (11)&446 (10)&1.6/4\\
   7&62  & \cite{na49a,na49b,na49c,na49d}
         &1.2 (0.2)&1.4 (0.3)&144 (3)&139 (3)&399 (13)&395 (10)&3.0/5\\
   7&7   & \cite{starbes}
         &1.0 (0.2)&1.5 (0.6)&147 (3)&138 (8)&424 (18)&368 (28)&8.0/4\\
   8&76  & \cite{na49b,na49c,na49d,na49e}
         &0.8 (0.1)&1.3 (0.4)&152 (3)&145 (5)&393 (15)&358 (18)&4.4/5\\
  11&5   & \cite{starbes}
         &1.0 (0.1)&1.9 (0.7)&157 (3)&142 (7)&310 (15)&278 (28)&0.8/4\\
  17&3   & \cite{na49c,na49d,na49e,na49f,na49g}
         &1.1 (0.2)&2.8 (0.4)&157 (3)&142 (3)&214 (14)&208 (8)&15/7\\
  39&    & \cite{starbes}
         &1.0 (0.2)&2.4 (0.8)&168 (4)&148 (8)&115 (13)& 98 (24)&1.2/4\\
  62&4   & \cite{stara,starb,starc}
         &1.3 (0.3)&2.3 (0.7)&169 (5)&155 (8)& 70 (20)& 65 (25)&8.0/7\\
 130&    & \cite{stard,starf,phenixa,phenixb}
         &1.6 (0.5)&2.5 (1.0)&169 (6)&157 (8)& 35 (23)& 25 (20)&4.4/5\\
 200&    & \cite{starg,starh,phenixc}
         &2.2 (0.4)&2.8 (0.8)&164 (3)&155 (6)& 31 (11)& 22 (16)&23/6\\
2700&    & \cite{lhc}
         &4.1 (0.6)&8.8 (0.8)&162 (3)&146 (3)& 14 (12)& -2  (7)&4.4/6\\
\hline
\end{tabular}
\end{center}
\caption{The freezeout parameters in 2CFO; the errors indicated are for
 single parameter variation.}
\label{tb.fit}\end{table*}

\begin{figure*}[tb]
\begin{center}
\includegraphics[scale=0.33]{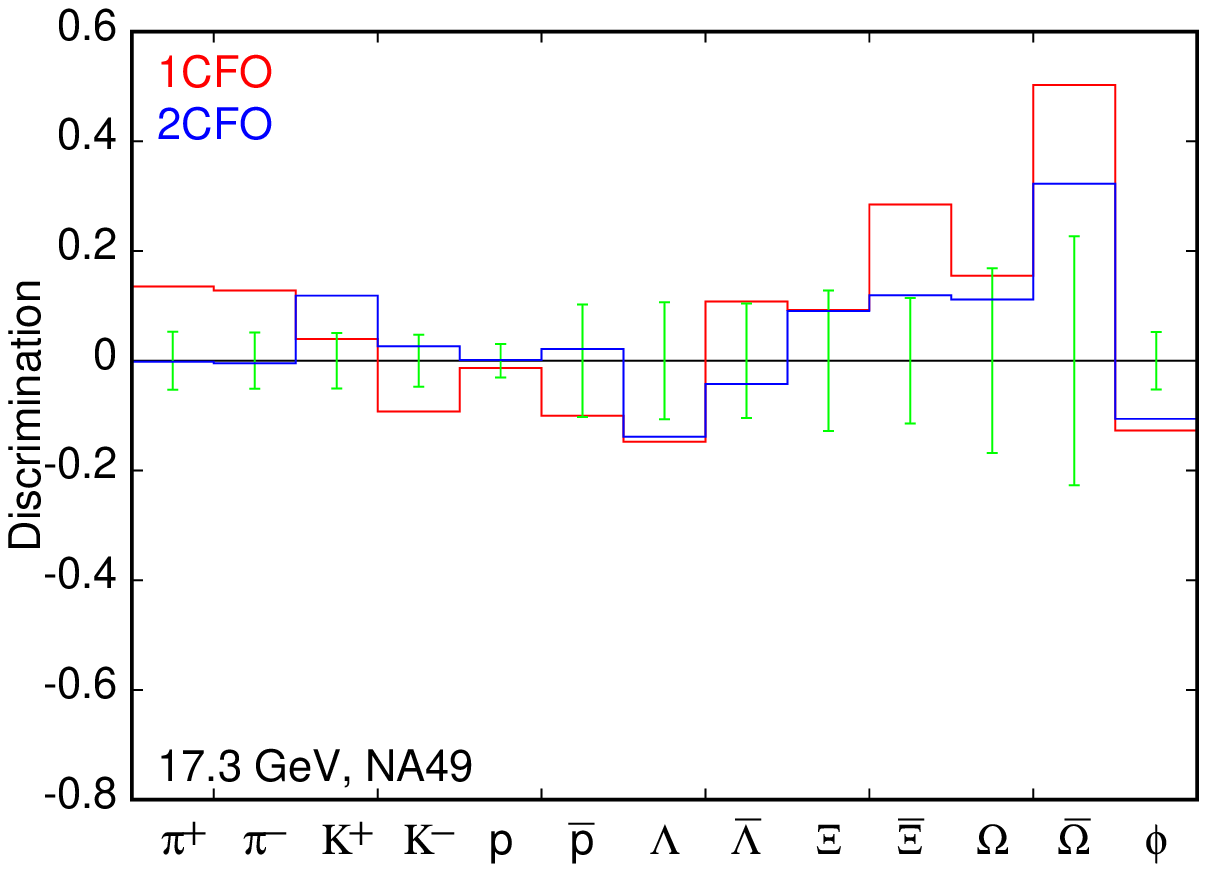}
\includegraphics[scale=0.33]{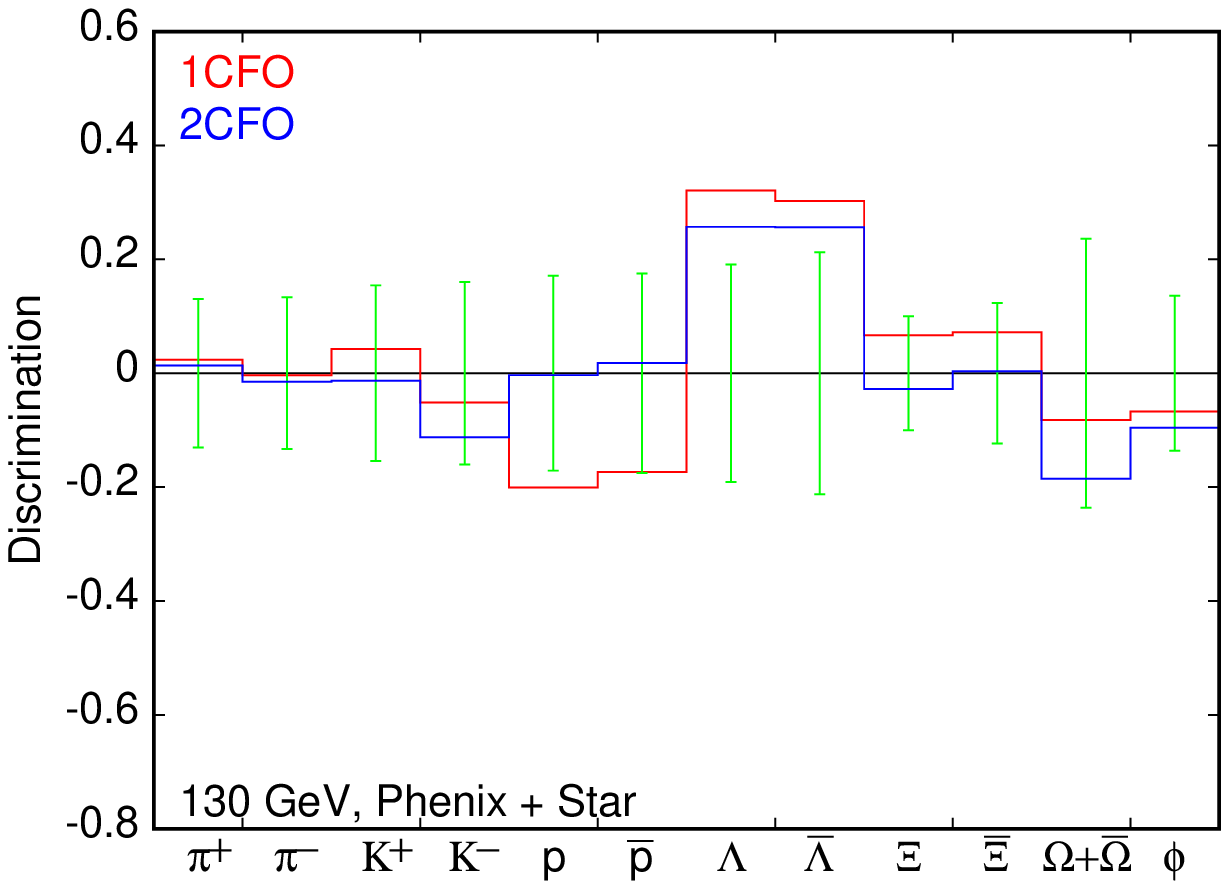}
\includegraphics[scale=0.33]{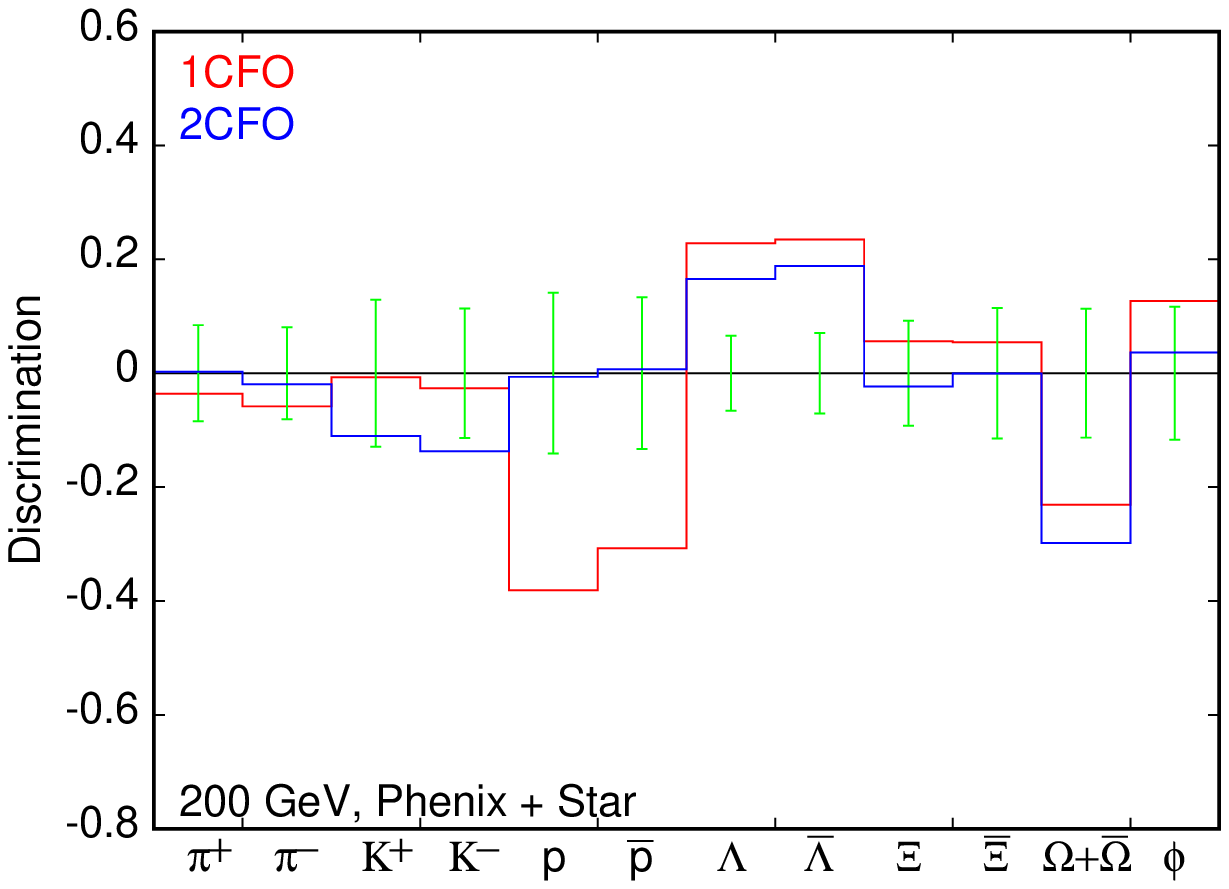}
\includegraphics[scale=0.33]{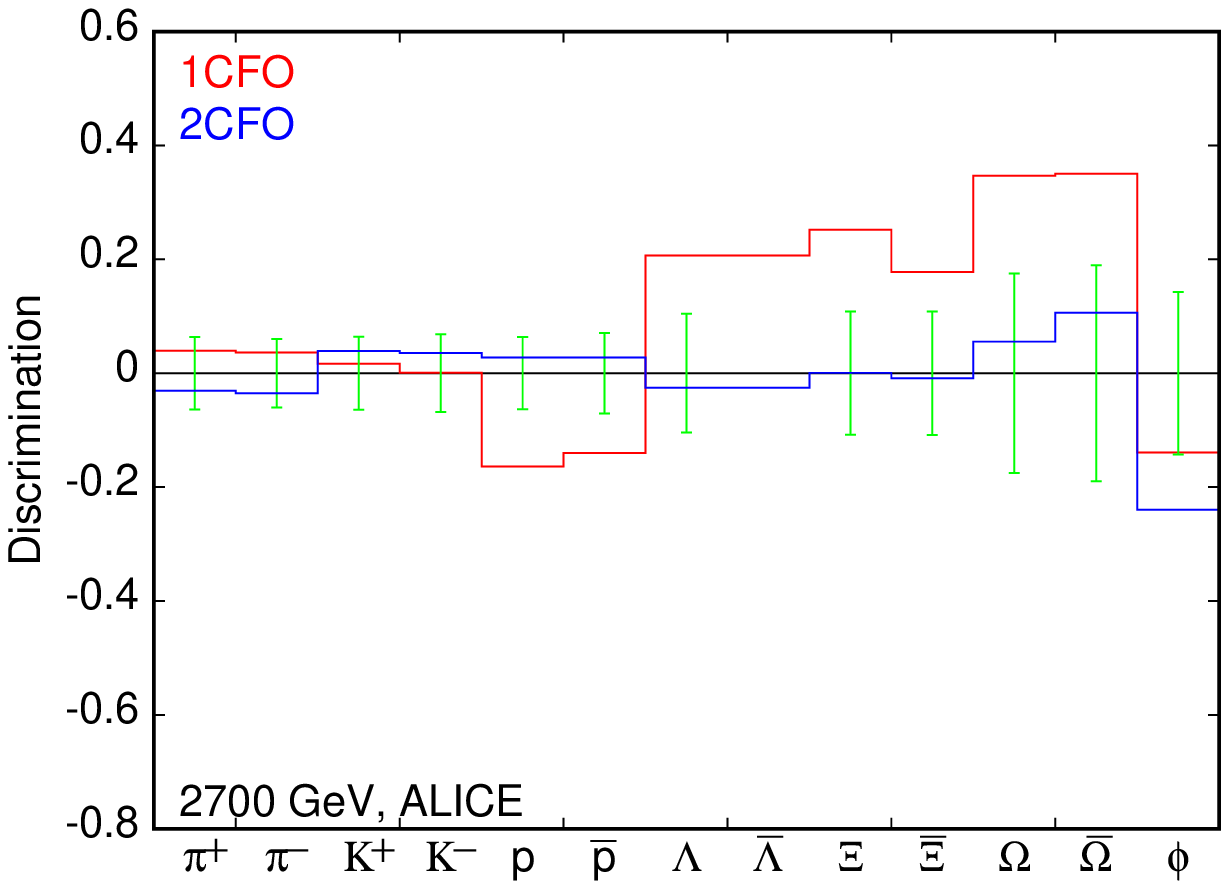}
\end{center}
\caption{Discrimination, $\Delta_h$, in different models of chemical freezeout,
 compared to relative error, $R_h$.}
\label{fg.discrimination}\end{figure*}

The paradigmatic case, which we call 1CFO, is merely the simplest of
a hierarchy of models because the kaons and pions happen to decouple
at the same time.  The next simplest possibility, which we call 2CFO,
is that all strange particles and $\phi$ decouple together at one time, and all 
other non-strange particles together at another time.  In this paper we examine
data on hadron yields \cite{na49a, na49b, na49c, na49d, na49e, na49f,
na49g, starbes, stara, starb, starc, stard, starf, phenixa, phenixb, starg,
starh, phenixc, lhc} within the framework of the second model to check
whether there is any evidence for a clear separation of pion and kaon
CFO. We report that there is indeed widespread evidence for 2CFO and
perhaps some evidence that even more detailed modelling of the chemical
kinetics is called for. Multiple freezeouts give us the ability
to use hadrons to look back into part of the history of the fireball. We
show here that interesting physics can follow from this.

Immediately after CFO, hadrons can be considered to be interacting
weakly. Then we can describe this matter as an ideal resonance gas while
preserving continuity of thermodynamic variables.  We populate this gas
with all hadron resonances with masses up to 2 GeV \cite{pdg}, neglect
widths and excluded volumes, set the strangeness undersaturation factor
{\gs} to unity, but treat the strange and electric charge fugacities
as in \cite{cleymans}. In this model for 1CFO we are able to treat the
data on yields at a level of precision similar to that achieved by more
detailed models: for example, for the 130 GeV data set from RHIC treated
in \cite{pbm} we find $\chi^2=12$, which is close to the quoted value
of $\chi^2=15$, with similar fitted parameters.  On the other hand, even
detailed 1CFO models are known to be unable to provide a description of
the data at $\sqrt S=2700$ GeV \cite{lhc}. Consequently, a late stage
non-thermal explanation has been put forward \cite{thlhc}. Although this
must certainly arise in a full kinetic computation, its strength must be
calibrated against the effect of multiple CFOs, since that too must occur.

In 2CFO, we fit yields to the same multi-component hadron resonance
gas but with one common temperature, \tns, baryon chemical potential,
\muns, and volume, \vns, for non-strange hadrons (except $\phi$), and a
different but common temperature, \ts, baryon chemical potential, \mus,
and volume, \vs, for strange hadrons and $\phi$.
Given a measurement of the yield, $Y_h$, of a hadron $h$, with
error, $\sigma_h$, and a model prediction, $Y_h^m$, the contribution
to $\chi^2$ from a specific hadron is
\beq
   \chi_h^2 = \left[\frac{\Delta_h}{R_h}\right]^2,
\label{channelchisq}\eeq
in terms of the discrimination, $\Delta_h=1-Y_h^m/Y_h$, and the
relative error, $R_h=\sigma_h/Y_h$.  It turns out that in most cases
$\chi^2_{\pi^\pm}$ and $\chi^2_{K^\pm}$ are small. Since this is
usually forced by small relative errors in these yields, they give
tight constraints on \vs, \vns, {\ts} and \tns. The baryon yields
determine {\mus} and \muns.  The best-fit parameters are shown in Table
\ref{tb.fit}.  In Figure \ref{fg.discrimination} we show $\Delta_h$
for the data sets which provide the most stringent tests of the models.
It is clear that 2CFO is a substantially better representation of the
data than 1CFO.

It is especially interesting to compare 1CFO and 2CFO at LHC in
Figure \ref{fg.discrimination}. There is clearly a tension between the
description of the strange and non-strange baryons in 1CFO, as a result
of which the $p/\pi$ ratios seemingly move away from equilibrium values.
2CFO resolves this tension with separate CFO in the two
sectors, as explained earlier. The figure shows that this resolves the
problem of some yields seeming to be non-thermal at the LHC. We note
that this tension is present, albeit at a statistically milder level,
also in the data taken at the other energies. There too the same mechanism
improves fits.

There is a smaller remnant problem with $\phi$, $\Lambda$, $\overline
\Lambda$, and $\overline \Omega$.  Since strangeness changing reactions
of baryons have high threshold energies, due to the large masses of
particles in the intermediate states, strange baryons could decouple
early.  A simple 3CFO model of this kind is sufficient to explain the
remaining discrepancies.  We are aware of the danger that this introduces
too many parameters, and therefore over-fits the data.  The exercise
is only meant to illustrate the fact that a detailed kinetic theory
computation may improve the description further.  A full computation
will also include the kinds of mechanisms explored in \cite{thlhc},
with the net result that the description will be finer.

\begin{figure}
\begin{center}
\includegraphics[scale=0.6]{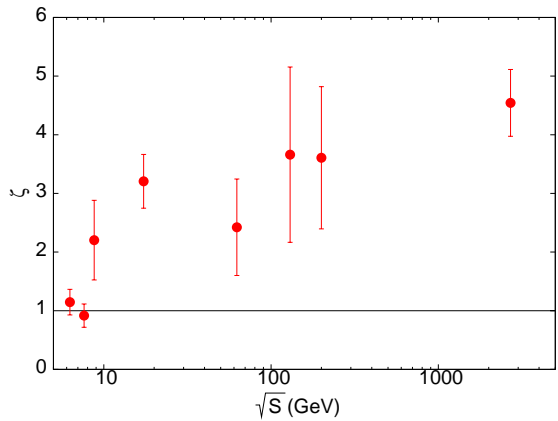}
\end{center}
\caption{The evolution of $\zeta=V_{NS}/V_c$ with $\sqrt S$.}
\label{fg.hbt}\end{figure}

We compare $V_{NS}$ with pion correlation volumes, $V_c$. Intensity
interferometry gives three measures of $V_c$, usually denoted by $R_l$,
$R_o$ and $R_s$ \cite{hbt}. We set $V_c=4\pi R_lR_oR_s/3$. Although
other definitions are possible, the differences are often not
statistically significantly.  At the two lowest energies the ratio
$\zeta=V_{NS}/V_c$ is around unity, and rises with $\sqrt S$, as shown in
Figure \ref{fg.hbt}. The rise can be attributed to the onset of collective
expansion. Interestingly, the evidence from $\zeta$ for further enhanced
collective flow at the LHC is weak, partly because of the large error
bars on $V_c$ at RHIC. Any improvement in the experimental errors
of the $\sqrt S$ dependence of $V_c$ would be very useful.

\begin{figure}
\begin{center}
\includegraphics[scale=0.6]{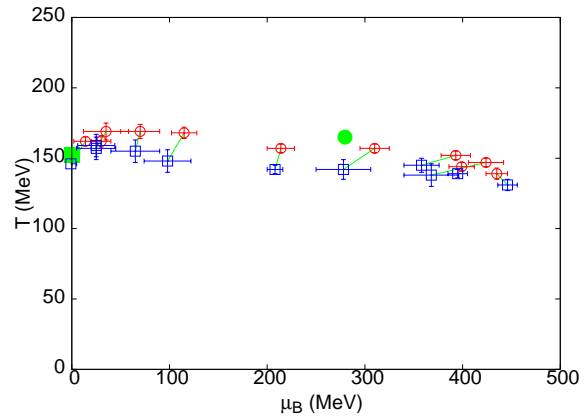}
\end{center}
\caption{The freezeout curves for 2CFO on the phase diagram (circles:
 strange, squares: non-strange). Points of equal $\sqrt S$ are joined by
 lines. The filled square is the predicted position of the QCD crossover
 at $\mu=0$ \cite{fodor} and the filled circle of the QCD critical
 point \cite{ilgti}.}
\label{fg.freeze}\end{figure}

The remaining parameters can be described by the fits
\beqa
\nonumber
  T_S    &=& (150\pm6) + \frac{130\pm40}L - \frac{295\pm60}{L^2},\\
\nonumber
  \mu_S &=&  -(40\pm40) + \frac{85\pm300}L + \frac{1900\pm500}{L^2},\\
\nonumber
  T_{NS} &=& (142\pm7)  + \frac{65\pm55}L - \frac{150\pm85}{L^2},\\
  \mu_{NS} &=&  -(85\pm20) + \frac{500\pm167}L + \frac{1000\pm250}{L^2},
\label{fits2fits}\eeqa
where $L=\log(\sqrt S/M_p)$ and $M_p$ is the proton mass. They are
plotted as freezeout curves in Figure \ref{fg.freeze}. At most energies
the two CFO surfaces are very close together. However, they move apart
at the LHC energy.

It is interesting to speculate on the reason for this.  One hypothesis
is that this is due to the significantly higher initial energy density,
so that an increased pion density delays its CFO. However an argument
against this is that the $K/\pi$ ratio remains constant from RHIC to LHC,
so the kaon density also rises in proportion, but the kaon CFO is not
delayed proportionately. A qualitatively different explanation relies
on the observation that between the two CFOs at very high $\sqrt S$,
the fireball passes close to the QCD chiral crossover point \cite{fodor},
which is in the critical region of the chiral transition in a theory with
$m_\pi=0$ \cite{karsch}. Chiral critical behaviour may then delay pion
CFO by lowering the scalar mass.  At lower energies, baryon chemical
potentials are larger, so the CFOs move away from the chiral critical
point.  However, data from higher energy runs at the LHC can crucially
test this explanation, as the system comes closer to the chiral critical
point with increasing $\sqrt S$.

Since the two CFOs at the same $\sqrt S$ lie on two different points of
the same trajectory of fireball evolution, there is more information
on late stage evolution than in 1CFO. The $K^+/\pi^+$ ratio peaks at
$\sqrt S=7.7$ GeV \cite{na49e}. The explanation \cite{horn} in 1CFO is
that this occurs at the energy where the fireball crosses from being
baryon dominated to meson dominated. This observation continues to
hold in 2CFO. However, the new ability to look back into the fireball
using different hadrons gives a new piece of information: that the
direction of the trajectory changes at this $\sqrt S$ (see Figure
\ref{fg.freeze}). This divergence of trajectories near the horn would
be interesting to understand from hydrodynamics or transport theory.

\begin{figure}
\begin{center}
\includegraphics[scale=0.6]{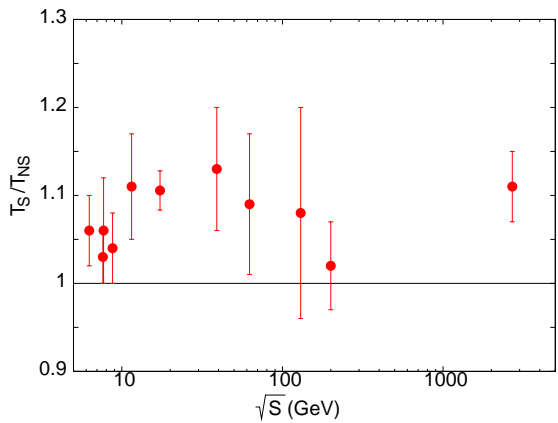}
\end{center}
\caption{The ratio $T_S/T_{NS}$ as a function of $\sqrt S$ possibly has
 a broad peak in the vicinity of $\sqrt S=17.3$ GeV.}
\label{fg.tratio}\end{figure}

The opposite seems to happen near the predicted position of the QCD
critical point \cite{ilgti}.  The comparison of the two trajectories
at $\sqrt S=11.3$ and 17.3 GeV (see Figure \ref{fg.freeze}) seems to
indicate a focusing of trajectories.  When trajectories pass near
a critical point, hydrodynamic anomalies of this kind are expected
\cite{srs}. Figure \ref{fg.tratio} shows that in this energy range
there is also some evidence of a broad peak in the ratio $T_s/T_{NS}$. A
second look at the freezeout curves in Figure \ref{fg.freeze} indicates
that this could be due to a delay in the CFOs. Correspondingly, the fits in
eq.\ (\ref{fits2fits}) indicate that both $T_S$ and $T_{NS}$ may reach a
minimum in this region. This possibly special behaviour in temperature
is not accompanied by any anomaly in $V_{NS}$, as seen from Figure
\ref{fg.hbt}. A putative explanation for these observations is a slow
expansion in out-of-equilibrium dynamics near the critical point, followed
by re-thermalization and delayed freezeout. These mild experimental hints
for interesting physics in this region could be critically tested if the
beam energy scan at RHIC runs long enough in the region of $\sqrt S$
between 19 and 11 GeV, and collects enough statistics to study hadron
yields more precisely.

In summary, we have shown that there is strong evidence for at least
two step CFO of hadrons in the fireball produced by the collision of
relativistic heavy-ions, and perhaps even some evidence of further
substructure. The separation between CFOs of different hadrons allows
us to look back into the fireball with strongly interacting probes
during the late stages of its evolution and learn about the physical
conditions there. Some of the exciting possibilities which arise are that
one may see some signals of the QCD chiral transition at LHC energies,
and of the QCD critical point at energies close to that where STAR sees
anomalies in the fluctuations of conserved quantities.  The LHC is now
inadvertently doing a energy scan; a similar very-high statistics scan
at the RHIC would be revealing.

We would like to thank Rajeev Bhalerao and Bedanga Mohanty for their
comments. RG wishes to thank the Department of Science and Technology,
Government  of India, for support under grant no. SR/S2/JCB-64/2007.
SG would like to thank the Department of Science and Technology,
Government  of India, for support under grant no. SR/S2/JCB-100/2011.

\end{document}